\documentclass[12pt,preprint]{aastex}
\shorttitle{Self-Induced Starburst at the Superbubble in M82}
\shortauthors{Matsushita et al.}

\begin{document}

\title{Starburst at the Expanding Molecular Superbubble in M82: \\
	Self-Induced Starburst at the Inner Edge of the Superbubble}

\author{Satoki Matsushita\altaffilmark{1}}
\altaffiltext{1}{Present address: Academia Sinica,
	Institute of Astronomy and Astrophysics,
	P.O.\ Box 23-141, Taipei 106, Taiwan, R.O.C.;
	satoki@asiaa.sinica.edu.tw}
\affil{Harvard-Smithsonian Center for Astrophysics,
	60 Garden St., MS-78, Cambridge, MA 02138}
\email{smatsushita@cfa.harvard.edu}
\author{Ryohei Kawabe}
\affil{National Astronomical Observatory of Japan,
	2-21-1 Osawa, Mitaka, Tokyo 181-8588, Japan}
\author{Kotaro Kohno}
\affil{Institute of Astronomy, University of Tokyo,
	2-21-1 Osawa, Mitaka, Tokyo 181-0015, Japan}
\author{Hironori Matsumoto and Takeshi G. Tsuru}
\affil{Department of Physics, Faculty of Science,
	Kyoto University, Sakyo-ku, Kyoto 606-8502, Japan}
\and
\author{Baltasar Vila-Vilar\'o}
\affil{Steward Observatory, University of Arizona, Tucson, AZ 85721}

\slugcomment{To be appeared in ApJ, 617, 20 December 2004 issue}

\begin{abstract}
We present high spatial resolution ($2\farcs3\times1\farcs9$ or 43 pc $\times$
36 pc at D = 3.9 Mpc) 100 GHz millimeter-wave continuum emission observations
with the Nobeyama Millimeter Array toward an expanding molecular superbubble
in the central region of M82.
The 100 GHz continuum image, which is dominated by free-free emission, revealed
that the four strongest peaks are concentrated at the inner edge of
the superbubble along the galactic disk.
The production rates of Lyman continuum photons calculated from 100 GHz
continuum flux at these peaks are an order of magnitude higher than those from
the most massive star forming regions in our Galaxy.
At these regions, high velocity ionized gas (traced by H41$\alpha$ and
[\ion{Ne}{2}]) can be seen, and H$_{2}$O and OH masers are also concentrated.
The center of the superbubble, on the other hand, is weak in molecular and
free-free emissions and strong in diffuse hard X-ray emission.
These observations suggest that a strong starburst produced energetic
explosions and resultant plasma and superbubble expansions, and induced
the present starburst regions traced by our 100 GHz continuum
observations at the inner edge of the molecular superbubble.
These results, therefore, provide the first clear evidence of self-induced
starburst in external galaxies.
Starburst at the center of the superbubble, on the other hand, begins to cease
because of a lack of molecular gas.
This kind of intense starburst seems to have occurred several times within
$10^{6}-10^{7}$ years in the central region of M82.
\end{abstract}


\section{INTRODUCTION}
\label{intro}

Messier 82 (NGC 3034, Cigar Galaxy) is known as one of the nearby
\citep[3.9 Mpc;][]{sak99}, bright irregular galaxies, and, therefore, has been
studied by many authors, at many wavelengths, and over many years.
Optical observations \citep{lyn63,sho98,ohy02} show remarkable features of
kpc-scale filaments extending above and below the disk of this galaxy.
Soft X-ray \citep*{wat84,kro85,bre95,tsur97,str97}, molecular gas
\citep*{nak87,mat00,tay01,wal02}, and dust \citep*{kun97,alt99} observations
also show filamentary structures outside the disk region.
These observations suggest that the filamentary structures are the result of
explosions, which occurred within the nuclear region.
The mass-to-luminosity ratio, $M/L$, of this galaxy is extremely low
($<0.02$, \citealp{rie79}; $\approx0.04$, \citealp{tel93}) as compared with
normal galaxies \citep[$5-10$,][]{rie78} and can be explained by bursts of
star formation or a starburst \citep{rie78,rie80}.

The starburst in M82 is believed to be triggered by a tidal interaction with
the nearby two-armed spiral galaxy M81.
Neutral hydrogen (\ion{H}{1}) emission observation toward the M81 group shows
that there is a tidal bridge between M81 and M82, suggesting that there is
a close encounter between these two galaxies $\sim10^{8}$ years ago
\citep*[e.g.,][]{cot77,yun94}.
Optical and near-infrared imaging and spectroscopic observations toward
the disk region ($0.5-1$ kpc outside from the galactic center) of M82 indicate
that the starburst occurred at this region $\sim10^{8}$ years ago, corresponds
to the episode of the tidal interaction with M81
\citep*{oco78,gal99,deg01,smi01}.
Since the present starburst region is located at the center, they suggested
that the global starburst in M82 has propagated inward from the disk to
the nuclear regions.
The evolution of starburst at the nuclear 500 pc region
is, on the other hand, still unclear.
Some authors suggest the nuclear starburst propagated outward
\citep[e.g.,][]{sat95,sat97,for03}, and some inward \citep[e.g.,][]{she95}.

Interferometric molecular gas observations show an arc-like deviation from
rigid rotation in position-velocity (PV) diagrams at $\sim140$ pc westward
from the galactic nucleus, and this deviation implies the existence of
an expanding molecular superbubble \citep{nei98,wei99,wil99}.
The existence of the superbubble was confirmed by imaging the shell-like
structure with the size of 260 pc $\times$ 170 pc \citep*{mat00,wil02}.
The age and the total energy of the superbubble are estimated to be about
$(1-2)\times10^{6}$ yr and $\sim(0.5-2)\times10^{55}$ ergs, which corresponds
to the total energy of $\sim10^{3}-10^{4}$ supernovae.
At the center of the superbubble, there is a ``2.2 $\mu$m secondary peak''
\citep[a primary peak corresponds to the nucleus;][]{die86,les90}, which is
dominated by luminous supergiants \citep*{joy87}.
Since luminous supergiants are the late phase of OB stars, it is reasonable to
suppose that a starburst has occurred at this near-infrared peak in the recent
past.
Actually, calculations of the initial mass function of this peak suggest that
the number of supernovae, which have already exploded within the age of
the superbubble, can explain the total energy of the superbubble.
It is therefore concluded that this superbubble has been produced as a result
of localized starburst at the ``2.2 $\micron$ secondary peak'' \citep{mat00}.

Detailed information about the nature of starburst in this region is, however,
still not clear, because the central region of M82 is deeply obscured
by dust even in the K-band (e.g., $A_{\rm K}\sim1.4$, \citealp{mcl93};
$A_{\rm K}\sim1.0$, \citealp{sat95}; $A_{\rm K}\lesssim0.8$, \citealp{tel91}:
These correspond to $A_{\rm V}\sim14$, $\sim10$, and $\lesssim8$,
respectively, if we assume $A_{\rm V}/A_{\rm K}=10$; \citealp{dra89}).
Hence it is very hard to derive detailed information --- such as the precise
flux of recombination lines --- in the optical or near-infrared.

On the other hand, 100 GHz continuum observations are generally dominated by
the free-free emission and would not be affected by the dust absorption, as is
the case for optical or near-infrared observations.
In fact, spectral energy distribution (SED) studies from the infrared to radio
wavelength region of M82 indicates that the 100 GHz continuum emission in M82
is dominated by free-free emission \citep{con92,mat98}.
A detailed SED study for small-scale structures toward the central region of
M82 by \citet{car91} based on their millimeter and centimeter interferometric
(BIMA and VLA) observations concluded that the free-free emission dominates
even in the small-scale structures of 100 GHz continuum emission
(i.e., 100 GHz continuum emission from all the peaks are dominated by free-free
 emission).
Hence, 100 GHz continuum observations can have great advantages for
the observation of the central starburst regions in M82, where a large amount
of dust exists.

In this paper, we report the structures and properties of the present starburst
regions and environment in M82 with our high spatial resolution 100 GHz
observations and compare them with previously published data.
We also discuss the triggering of new starbursts by the past energetic
starburst and the resultant superbubble expansion, namely, self-induced
starburst.

\section{OBSERVATIONS}
\label{obs}

We obtained aperture synthesis 100 GHz continuum images toward the central
region of M82 with the Nobeyama Millimeter Array during 1997 December -- 1998
April.
Three (AB, C, and D) configurations of six 10 m antennas, equipped with
tunerless SIS receivers \citep*{sun94}, were used for the observations.
As back-end, we used an XF-type spectro-correlator Ultra Wide Band Correlator
\citep[UWBC;][]{oku00}, which has a 1 GHz bandwidth and a 6 GHz intermediate
frequency.
Since the UWBC can process both the upper side band (USB) and the lower side
band (LSB), our continuum data were obtained from the USB of the HCN(1 -- 0)
observations (the rest frequency of the HCN J = 1 -- 0 line is 88.632 GHz, and
the observed frequency is 88.884 GHz).
Hence the central frequency of the continuum observations corresponds to
100.884 GHz.
The band-pass calibration was done with 3C273, and 0923+392 was observed every
10 minutes as a phase and amplitude calibrator.
The flux scale of 0923+392 was determined by comparisons with Mars and Uranus.
The uncertainty in the absolute flux scale is estimated to be $\sim$ 20\%.

The data were reduced using the NRO software package ``UVPROC II''
\citep*{tsut97}, and the final maps were made and CLEANed with the NRAO
software AIPS.
We analyzed the 100 GHz continuum data with uniform $uv$ weighting at
the CLEAN process, and the resultant resolution was
$2\farcs3\times1\farcs9$ (43 pc $\times$ 36 pc at D = 3.9 Mpc).
We note that this linear scale of the beam size is similar to
the typical diameter of a giant molecular cloud (GMC).

\section{RESULTS}
\label{res}

\subsection{Distribution of 100 GHz Continuum Emission}
\label{res-dist}

\begin{figure}
\plotone{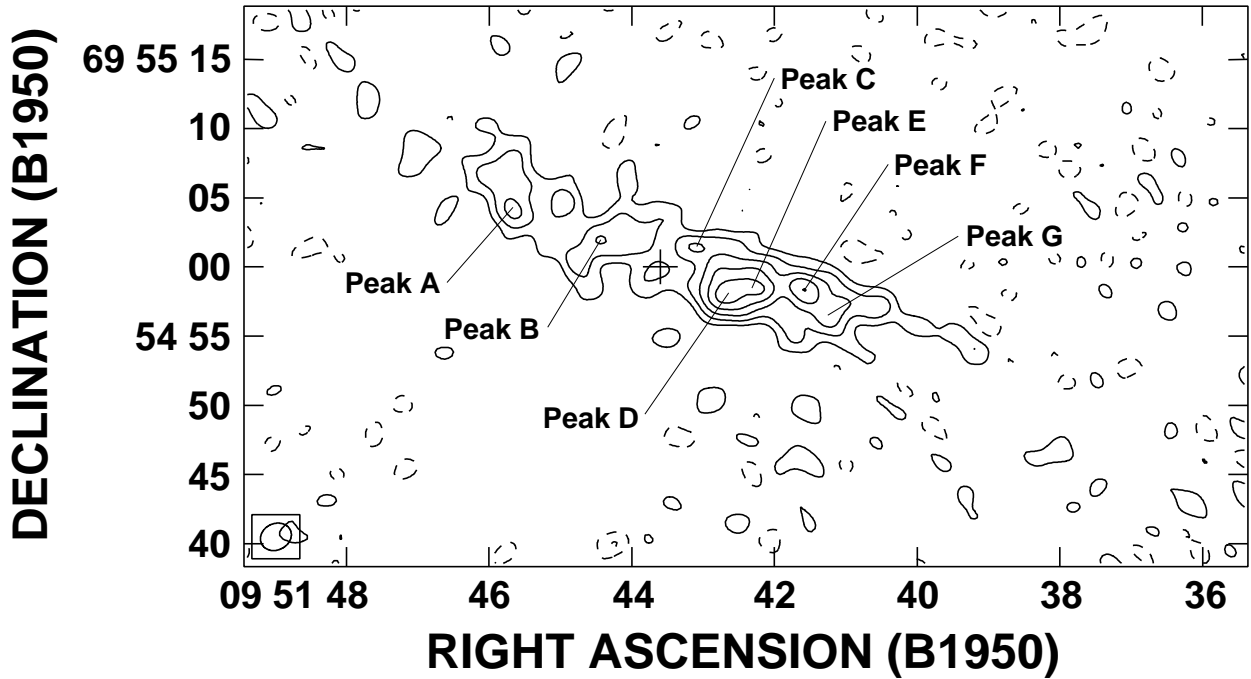}
\figcaption{
	Uniform $uv$ weighting 100 GHz continuum image of the central region
	of M82.
	The contour levels of the map are
	$-4, -2, 2, 4, 6, 8,$ and $10\sigma$,
	where $1\sigma$ = 2.0 mJy beam$^{-1}$.
	The synthesized beam ($2\farcs3\times1\farcs9$ or 43 pc $\times$ 36 pc)
	is shown at the bottom-left corner.
	The plus mark indicates the position of the galactic nucleus
	determined from the peak of the strongest 2.2 $\micron$ source of
	$\alpha$(B1950)=$9^{\rm h}51^{\rm m}43\fs6$ and
	$\delta$(B1950)=$69^{\circ}55'00''$ \citep{les90}.
	Peaks A-G are labeled in the image.
\label{fig-uni-cont}}
\end{figure}

The 100 GHz continuum map made with uniform $uv$ weighting is shown in
Figure~\ref{fig-uni-cont}.
We defined a ``peak'' as
(1) detected in the map with $6\sigma$ or more and
(2) separated from other peaks by more than one beam width.
We found seven peaks in the map (Peaks A--G in the figure), and all peaks were
aligned along the galactic disk (position angle or P.A.\ of 76\arcdeg).
The four strongest peaks, D--G, correspond to the two strongest peaks in
the previously published lower spatial resolution 100 GHz continuum images
\citep{car91,bro93,sea96,nei98}.
Other weaker peaks in our map correspond to those in the previously published
images.
Note that our lower resolution ($3\farcs7\times3\farcs3$) and higher
signal-to-noise ratio map
(i.e., natural $uv$ weighting map; not shown in this paper)
is consistent with other published results.

Since the uniform $uv$ weighting increases the noise level of the map,
the diffuse extended emission may be masked by the noise, and the total flux
of the 100 GHz continuum emission may decrease compared with the natural $uv$
weighting image.
However, the total flux of the uniform $uv$ weighting map is, in fact,
$\sim$ 80\% of that of the natural $uv$ weighting map, which suggests that
most of the flux is not missing.
On the other hand, the total flux of our natural $uv$ weighting map is
0.43 Jy, and the missing flux compared with the single-dish flux
\citep*[$\sim$ 0.54 Jy;][]{jur78} is $\sim$ 20\%, which is almost comparable
with the uncertainty in the absolute flux scale.
We therefore did not recover the missing flux and conclude that the high
spatial resolution starburst properties, as reported here, are accurate at
the 30\% level.

\begin{figure}
\plotone{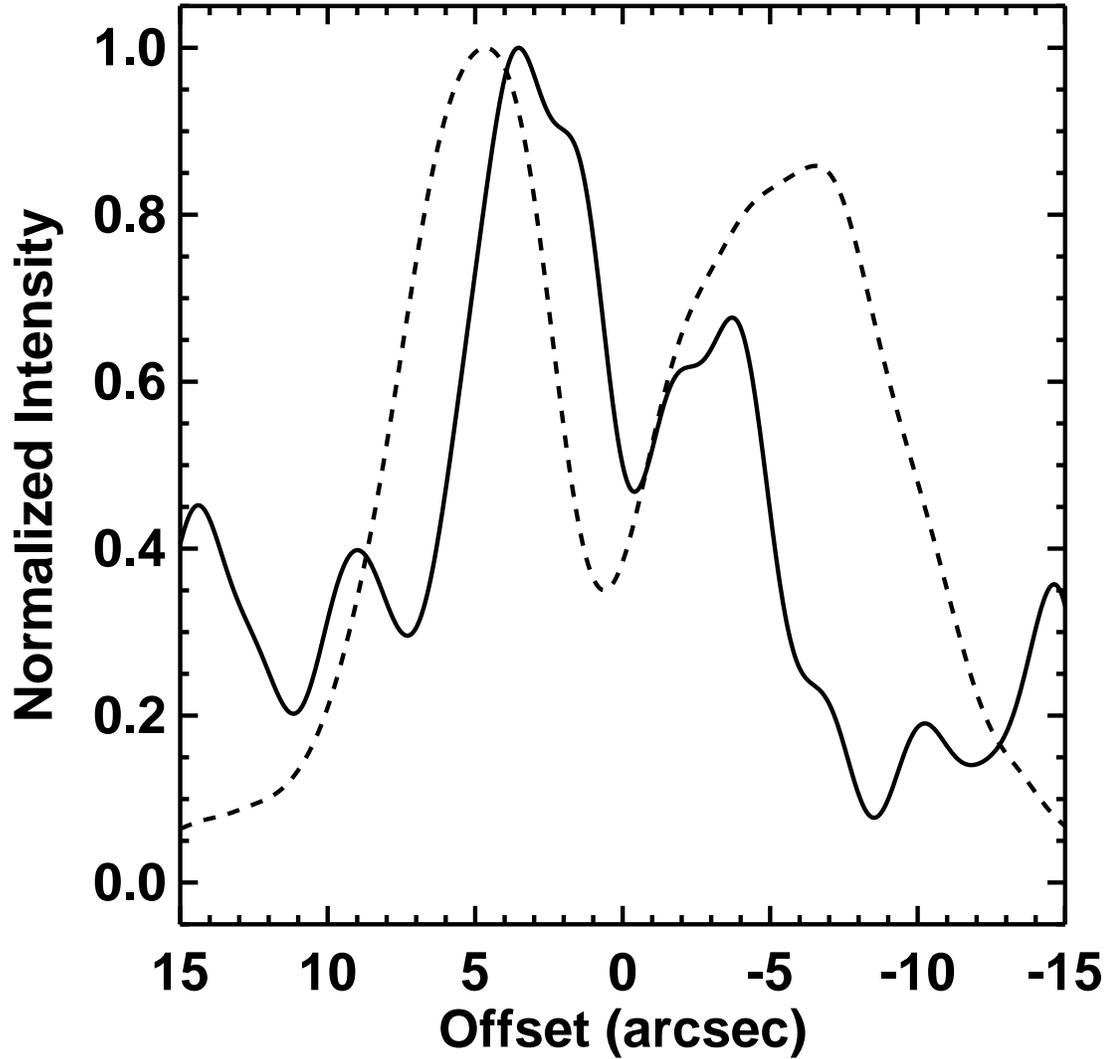}
\figcaption{
	100 GHz continuum (solid line) and $^{12}$CO(1 -- 0) (dashed line)
	intensity plots of the superbubble sliced along the major axis
	(P.A.\ = 76\arcdeg), namely along the disk, of the galaxy.
	The vertical axis is normalized intensity, and the horizontal axis is
	offset from the center of the molecular superbubble, which is
	$\alpha$(B1950)=$9^{\rm h}51^{\rm m}41\fs9$ and
	$\delta$(B1950)=$69\arcdeg54\arcmin57\farcs6$.
\label{fig-slice}}
\end{figure}

Figure~\ref{fig-cont-co} shows the $^{12}$CO(1 -- 0) superbubble image
\citep[contour map;][]{mat00} overlaid on the 100 GHz continuum image
(colorscale map).
The dashed line shows the possible oval structure of the superbubble.
This figure shows that {\it the four brightest peaks (D-G) in the 100 GHz
continuum image are located at the inner edge of the molecular superbubble
along the galactic disk of M82.}
We also made an 100 GHz continuum intensity plot of the superbubble sliced
along the major axis (P.A.\ = 76\arcdeg), namely along the galactic disk, of
the galaxy to compare with that of the $^{12}$CO emission
(Fig.~\ref{fig-slice}).
The continuum emission has a double-peaked structure in the plot:
The intensity is weak at the center of the superbubble (around zero offset in
the figure) and has peaks at both sides of the bubble center, which are
located around $\pm(2\arcsec-4\arcsec)$ offsets.
The plot of the $^{12}$CO emission also shows a double-peaked structure.
The emission is weak at the center of the bubble, but peaks are located
around $\pm(4\arcsec-7\arcsec)$, outside the peaks of the continuum emission.
This figure demonstrates that the 100 GHz emission is located at the inner edge
of the molecular superbubble.

Since the 100 GHz continuum emission in M82 is dominated by free-free emission,
even in small-scale structures (see \S\ref{intro}), these 100 GHz continuum
emitting regions represent starburst regions.
This distribution, therefore, strongly suggests that the present starburst
regions in M82 are concentrated at the inner edge of the molecular superbubble.

\subsection{Production Rates of Lyman Continuum Photons}
\label{res-pro}

\begin{figure}
\plotone{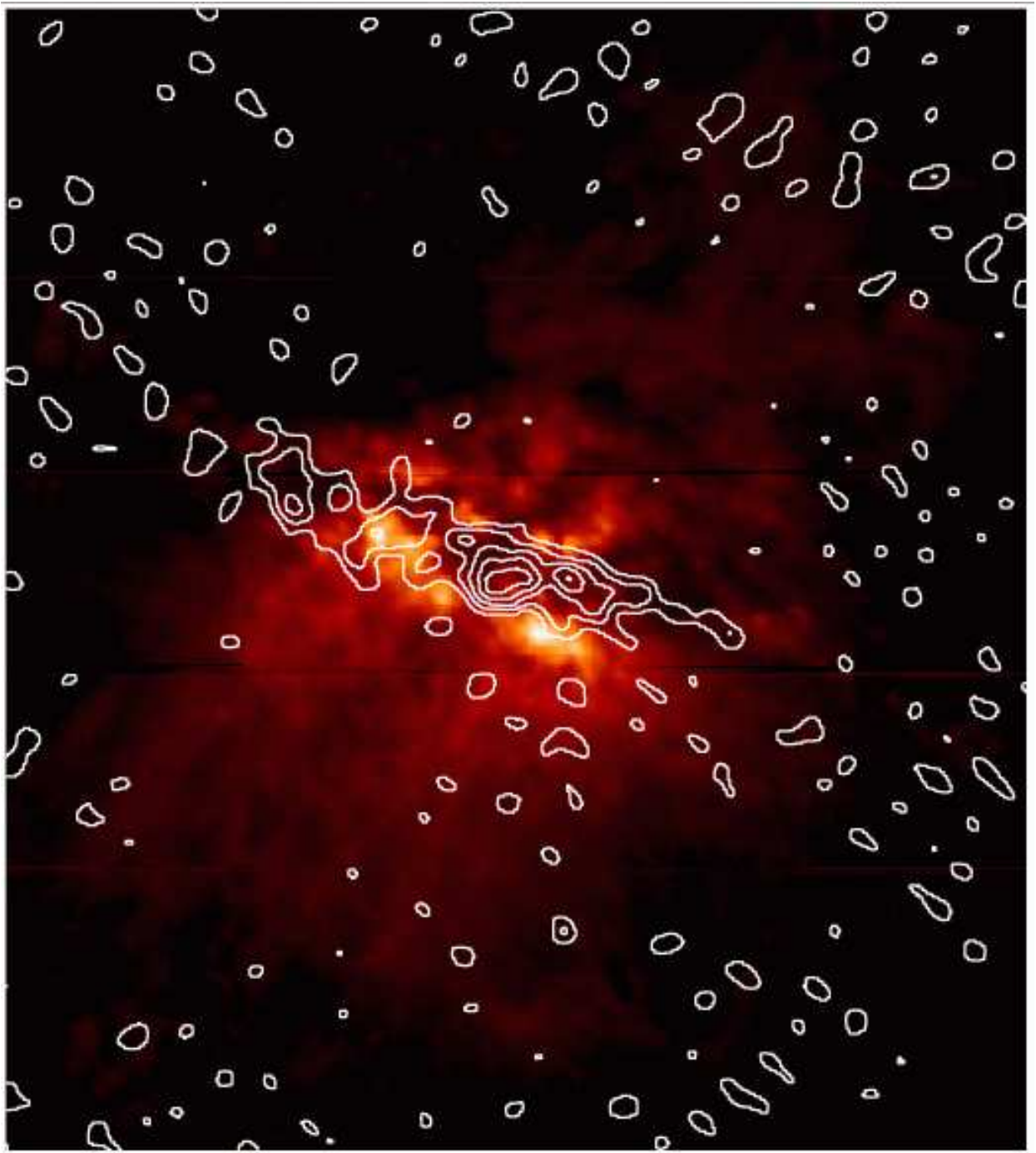}
\figcaption{
	H$\alpha$ narrow band image taken by the Subaru 8~m telescope
	\citep[colorscale map;][]{ohy02} overlaid on the 100 GHz
	continuum image (contour map).
	The contour scale is the same as in Fig.~\ref{fig-uni-cont}.
\label{fig-cont-ha}}
\end{figure}

The 100 GHz continuum emission in M82 is dominated by free-free emission,
as mentioned above, so that the strength of the emission is directly related
to star formation activities like H$\alpha$ emission at optical wavelength.
Furthermore, 100 GHz continuum emission does not be affected by the dust
absorption like the H$\alpha$ emission.
Figure~\ref{fig-cont-ha} shows the H$\alpha$ image
\citep[colroscale map;][]{ohy02} overlaid on the 100 GHz continuum image
(contour map).
The map clearly shows that the brightest 100 GHz continuum peaks are
totally obscured by dust in the H$\alpha$ image, indicating that 100 GHz
continuum observation is a strong tool to image the active starburst regions
directly.

We therefore use the 100 GHz continuum emission flux (i.e., absorption-free
free-free emission flux), $S_{\rm T}$, to estimate the production rates
of Lyman continuum photons, $N_{\rm Ly}$.
The absorption-free $N_{\rm Ly}$ can be calculated as
\begin{equation}
\left(\frac{N_{\rm Ly}}{[{\rm s}^{-1}]}\right)
	\sim1.8\times10^{51}
	\left(\frac{T_{\rm e}}{10^{4}~{\rm [K]}}\right)^{-0.45}
	\left(\frac{\nu}{{\rm 100.88~[GHz]}}\right)^{0.1}
	\left(\frac{D}{{\rm 3.9~[Mpc]}}\right)^{2}
	\left(\frac{S_{\rm T}}{{\rm [mJy]}}\right),
\label{eq-Nly}
\end{equation}
where $T_{\rm e}$ and $\nu$ are the electron temperature and the observation
frequency, respectively \citep{rub68,con92}.
We adopted $D=3.9$ Mpc \citep{sak99} for the distance of M82.
Since radio recombination line observations toward the central region of M82
suggest that $T_{\rm e}$ of the ionized gas (different from hot plasma which
can be seen in X-ray emission) ranges between $\sim$5,000 K and $\sim$10,000 K
\citep*{sea96,sea94,pux89}, we used $T_{\rm e}$ = 5,000 K and 10,000 K for
the calculations.

We substitute the peak flux density of the 100 GHz continuum averaged over
the beam size of $2\farcs3\times1\farcs9$ (43 pc $\times$ 36 pc) into
eq.~(\ref{eq-Nly}), and derived $N_{\rm Ly}$ at the specific regions
(Peaks A-G in Fig.~\ref{fig-uni-cont}) in M82 (Table~\ref{tab1}).
We find that $N_{\rm Ly}$ in the present star forming regions averaged over
the linear scale of about $30-40$ pc are as high as
$\sim(3-6)\times10^{52}$ photons s$^{-1}$ for the four strong peaks (D--G;
the western side of the nucleus) and $\sim(2-3)\times10^{52}$ photons s$^{-1}$
for other weaker peaks (the eastern side of the nucleus).
The uncertainty of $N_{\rm Ly}$ is about
$(0.4-0.5)\times10^{52}$ photons s$^{-1}$, which corresponds to $1\sigma$
statistical uncertainty in the continuum map.

To find out the validity of our estimation of $N_{\rm Ly}$, we compared our
estimation with those of infrared observations.
\citet{for01} estimated $2\farcs25\times2\farcs25$ aperture size $N_{\rm Ly}$
using their Br$\gamma$ image, which is corrected for extinction of
$A_{\rm V}\sim45\pm20$.
The $N_{\rm Ly}$ at their position B2, which corresponds to our Peak D,
is $(3.24\pm1.33)\times10^{52}$ photons s$^{-1}$, consistent with that of
Peak D of $(4.2-5.7)\times10^{52}$ photons s$^{-1}$ (Table~\ref{tab1}).
\citet{sat97} also estimated $N_{\rm Ly}$ using their Br$\gamma$ image,
corrected for reddening of about $A_{\rm V}\sim10$.
They found 12 point-like sources at the central 500 pc region using
an extinction corrected K-band image, and calculated for $N_{\rm Ly}$ with
$2''$ aperture size, similar to our beam size.
$N_{\rm Ly}$ values for the eastern side of the nucleus show
$\sim(0.3-0.8)\times10^{52}$, and those of western side show
$\sim(0.8-1.4)\times10^{52}$.
The tendency that $N_{\rm Ly}$ of the western side is about twice as
strong as that of the eastern side is the same as our results.
Their absolute values are, however, a factor of a few smaller than our
estimation and that of \citet{for01}.
This inconsistency maybe due to their smaller reddening correction.
In summary, our $N_{\rm Ly}$ estimation using 100 GHz continuum flux and
that using infrared emission line data are consistent each other.

\section{DISCUSSION}
\label{dis}

In the previous section, we show that the present starburst regions are
strongly concentrated at the inner edge of the expanding molecular
superbubble.
In this section, we discuss the starburst activities, environment, and
history around the superbubble.

\subsection{Comparison of Star Formation Activity with Our Galaxy and Nearby Galaxies}
\label{dis-comp}

Since the linear scale of our beam size is comparable to those of GMCs, it is
worth comparing the star forming activity around the superbubble in M82 with
those in active star forming GMCs in our Galaxy and nearby galaxies using
$N_{\rm Ly}$.
Free-free continuum emission is, however, optically thin at 100 GHz, so there
may exist some overlap of the starburst regions, since M82 is an edge-on
galaxy.
We therefore evaluated the effect of overlapping based on the structure of
the superbubble and estimated that a few GMCs can overlap at Peaks D and G,
since the shell structure is parallel to the line-of-sight.
However, there may be none at Peaks E and F, which are close to the bubble
center (i.e., the shell structure is perpendicular to the line-of-sight).
There is also a possibility that starburst regions around the superbubble
overlap with starburst regions in the background disk.
This possibility is, however, very small, because stars are formed from
molecular gas, and almost all the molecular gas toward the superbubble is
dynamically associated with it \citep{nei98,wei99,wil99,mat00}.
In addition, radio recombination lines, which also trace star forming regions,
toward the superbubble are also dynamically associated with it
(see \S\ref{dis-inducesb-ion}).
These discussions indicate that Peaks E and F are starburst regions at
the inner edge of the superbubble with no overlap, and we therefore compare
$N_{\rm Ly}$ at Peaks E and F with those at active star forming GMCs in our
Galaxy and nearby galaxies.

At Peaks E and F, $N_{\rm Ly}$ are $5.2\times10^{52}$ and $5.0\times10^{52}$
photons s$^{-1}$ for $T_{\rm e}$ = 5,000 K and $3.8\times10^{52}$ and
$3.6\times10^{52}$ photons s$^{-1}$ for $T_{\rm e}$ = 10,000 K, respectively
(see Table~\ref{tab1}).
These $N_{\rm Ly}$ are an order of magnitude higher than those of active star
forming regions in our Galaxy:
The most massive and luminous optically visible giant \ion{H}{2} region
NGC 3603 \citep{gos69}, which is $\sim10$ kpc away from the Galactic Center,
has $N_{\rm Ly}$ of $1.1\times10^{51}$ photons s$^{-1}$ \citep*{smi78}.
The \ion{H}{2} region complex G$49.5-0.4$ in one of the most luminous star
forming regions W51, which is 7.6 kpc away from the Galactic Center
\citep{smi78}, has $N_{\rm Ly}$ of $8.9\times10^{50}$ photons s$^{-1}$
\citep{osi00}.
The diameters of these regions are about 50 pc and 30 pc, respectively
\citep{osi00},
which are almost the same linear scale as Peaks E and F.
This comparison indicates that the ongoing starburst regions in M82 are at
least 30 times more active in terms of ionizing flux than the active star
forming regions in our Galaxy.

Millimeter-wave (3.4 mm or 88 GHz) continuum image toward the center of
the nearby spiral galaxy IC 342 show a continuum peak at a star forming GMC
\citep[cloud B in][]{dow92}.
The peak intensity is $3.3\pm0.6$ mJy beam$^{-1}$ with the beam size of
$2\farcs7\times2\farcs7$ \citep[24 pc $\times$ 24 pc at D = 1.8 Mpc;][]{mcc89},
and seems to be dominated by free-free emission.
Substituting these values in Eq.~\ref{eq-Nly}, $N_{\rm Ly}$ can be calculated
as $(1.7\pm0.3)\times10^{51}$ photons s$^{-1}$ for $T_{\rm e}$ = 5,000 K and
$(1.2\pm0.2)\times10^{51}$ photons s$^{-1}$ for $T_{\rm e}$ = 10,000 K.
H$\alpha$ and Pa$\alpha$ emission line observations toward the central $3'-4'$
region of grand-design spiral galaxy M51 using the WFPC2 and NICMOS instruments
on the Hubble Space Telescope depicted more than 1000 \ion{H}{2} regions with
their typical sizes of $10-100$ pc \citep{sco01}.
The fraction of the sizes of \ion{H}{2} regions peak around $20-40$ pc, which
is independent of their locations (including nuclear, arm, and interarm
regions).
The $N_{\rm Ly}$ of the \ion{H}{2} regions with the sizes less than 50 pc
show $\sim7\times10^{50}$ photons s$^{-1}$.
In both galaxies, star formation activities in terms of ionizing flux with
the scale of $30-40$ pc show an order of magnitude lower than the starburst
GMCs in M82.

Comparisons between starburst GMCs in M82 and star forming GMCs in our Galaxy,
M51, and IC 342 suggest that the star formation in M82 is quantitatively
different from those in our Galaxy, M51 and IC 342, namely, M82 has an order
of magnitude higher star formation efficiency in GMC scale.
Recent high angular resolution optical and near infrared observations toward
nearby galaxies show many super star clusters (SSCs) with their size at most
10 pc.
Our observations are limited by the beam size of $30-40$ pc, thus we cannot
compare them directly.
It will be interesting to perform sub-arcsec resolution ($<20$ pc) 100 GHz
continuum observations toward these starburst regions in M82 to resolve
SSCs and compare with those of other galaxies in the future.
These observations may help to understand what causes the qualitative
differences between starburst GMCs in M82 and star forming GMCs in our
Galaxy and nearby galaxies.

\subsection{Stellar Compositions in the Ongoing Starburst Regions}
\label{dis-mf}

In the previous section, we estimated the production rates of Lyman continuum
photons ($N_{\rm Ly}$) from the ongoing starburst regions in M82.
Based on this information, it is possible to estimate the stellar compositions
in the starburst regions by assuming a mass function (MF).

\subsubsection{Mass Function Calculations}
\label{dis-mf-calc}

We adopted an extended Millar-Scalo MF of $dN/dm\propto m^{-2.5}$ with lower
and upper mass limits of 1 and 100 M$_{\odot}$ \citep{ken83}.
To estimate the total $N_{\rm Ly}$ from all the stars calculated from the MF,
the value of $N_{\rm Ly}$ from stars with each stellar mass is needed.
We used the stellar model of \citet*{vac96}, who calculated $N_{\rm Ly}$ for
the O-type and early B-type stars, and their evolutionary mass model (stellar
mass derived from an evolutionary track) for the luminosity class V (main
sequence).
The reason for using this class is the following:
Since the ongoing starburst regions are located just at the inner edge of
the superbubble, and these seem to be induced by the past starburst that made
the superbubble (see \S\ref{dis-inducesb}),
it is reasonable to suppose that the present starburst regions are younger than
the expanding timescale of the superbubble.
The age of the superbubble is about $(1-2)\times10^{6}$ years
\citep{wei99,mat00,wil02}; hence the starburst should have occurred on less
than this timescale, which is similar to or less than the lifetime of
early-type stars.
It is therefore appropriate to assume that all the stars in the present
starburst regions are still on the main sequence.

Since the calculations of the stellar model in \citet{vac96} are done with each
spectral type, we interpolated their $N_{\rm Ly}$ for the smaller interval of
the stellar mass with fourth-order polynomial fitting.
The fitting residual is 1.2\%, which is much smaller than other errors
(e.g., statistical noise and absolute flux error of our observations).
The extrapolations for very massive stars and for late-type stars (later than
late B-type stars) were done with the same function as the interpolation.
Using this ``interpolated/extrapolated $N_{\rm Ly}$ - stellar mass'' relation
and the MF explained above, we calculated the stellar compositions for each
starburst region (Peaks A-G).
The calculated stellar numbers with mass larger than 8 M$_{\odot}$, which are
the progenitors of supernovae, are summarized in Table~\ref{tab1}.

We also estimated the dependence of MF slopes and stellar mass limits.
If we apply a steeper MF of $dN/dm\propto m^{-3.0}$, which is close to
the slope of Miller-Scalo solar neighborhood MF of -3.3 \citep{mil79}, 
the calculated $>8$ M$_{\odot}$ stellar numbers will increase about a factor
of 2.
On the other hand, if we apply a shallower MF of $dN/dm\propto m^{-2.0}$,
the calculated $>8$ M$_{\odot}$ stellar numbers will decrease about a factor
of 2.
If we lower the upper mass limit to 50 M$_{\odot}$, the calculated
$>8$ M$_{\odot}$ stellar numbers will increase about a factor of 2.
Changing the lower mass limit affects to the calculations only slightly, since
low mass stars emit small amounts of Lyman continuum photons.
\citet{rie93} computed the starburst model for M82, and concluded that
the MF model of $dN/dm\propto m^{-3.0}$ with the upper mass limit of
80 M$_{\odot}$ will fit their observations the best.
\citet{for03} also performed starburst model calculations, and found out that
their data are consistent with the MF model of $dN/dm\propto m^{-2.35}$ with
the upper mass limit of $\gtrsim50-100$ M$_{\odot}$.
In either case, the difference between our and their MF models will not
produce an order of magnitude difference in the calculated $>8$ M$_{\odot}$
stellar numbers.

There is also a possibility that the superbubble expansion timescale is longer
than the previous estimation of $(1-2)\times10^{6}$ years.
This timescale is calculated with
[the size of the superbubble]/[expansion velocity], but if the acceleration
of the molecular gas to the present expansion velocity is gradual
(i.e., if the expansion velocity is slow when the superbubble is small),
the expansion timescale should be longer than the estimation.
If this is the case, some of the highest mass stars should have already been
exploded by supernovae, and this effect should be considered in the MF
calculations.
This effect corresponds to lowering the upper mass limit of the MF
calculations and, therefore, increasing the $>8$ M$_{\odot}$ stellar numbers.
The superbubble expansion timescale shorter than $(1-2)\times10^{6}$ years
can also be possible, due to the deceleration of the expanding velocity
\citep{wei99,gar01}.
In this case, the timescale is shorter than the age of the high-mass (OB type)
stars, and we can assume that there are no supernova explosions in the past.
Therefore, this case does not affect any MF calculations.

\subsubsection{Estimated Supernova Energy Release from the Ongoing Starburst
	Regions}
\label{dis-mf-sn}

It is interesting to find out whether the present starburst is more active
than the past starburst that made the molecular superbubble.
In order to compare the energetics of the present starburst with that of
the molecular superbubble, we calculated $>30$ M$_{\odot}$ stars, whose
lifetime corresponds to the dynamical timescale of the superbubble of
$10^{6}$ years, at Peaks E and F.
We found $\sim(12\pm1)\times10^{2}$ stars for $T_{\rm e}$ = 5,000 K and
$\sim(9\pm1)\times10^{2}$ stars for $T_{\rm e}$ = 10,000 K, which corresponds
to about $10^{3}$ supernova explosions with a total energy release of
$\sim10^{54}$ ergs.

The efficiency of the supernova explosion energy input to the bubbles/chimneys
varies.
A single supernova explosion releases its energy to the surrounding ISM with
an efficiency of $3-8$\%, depending on the density of the ISM \citep{che74}.
Multiple supernova explosions, on the other hand, release their energy into
the ISM more efficiently, with up to 20\% of their total energy \citep{mcc87}.
If we take the efficiency of $3-20\%$, bubbles/chimneys with
energy of $10^{52-53}$ ergs will be produced at Peaks E and F in the future,
which is similar to the observed bubbles in our Galaxy or nearby galaxies
\citep[e.g.,][]{ten88,mar98}.
This energy is, however, a few orders of magnitude lower than that of
the molecular superbubble in M82
\citep[$\sim(0.5-2)\times10^{55}$ ergs;][]{mat00}
and suggests that the present starburst regions are less active than
the starburst that made the molecular superbubble.

This possibility may not be the case if the expansion timescale of
the superbubble is longer than expected.
The longer timescale increases the energy release from the present starburst
regions owing to the increase in the number of supernovae
(see \S\ref{dis-mf-calc}).
The energy input into the surrounding ISM, therefore, would be closer to
the energy of the present superbubble.

\subsection{Induced Starburst at the Inner-Edge of the Superbubble}
\label{dis-inducesb}

\subsubsection{Structure of the Superbubble}
\label{dis-inducesb-struct}

\begin{figure}
\epsscale{0.7}
\plotone{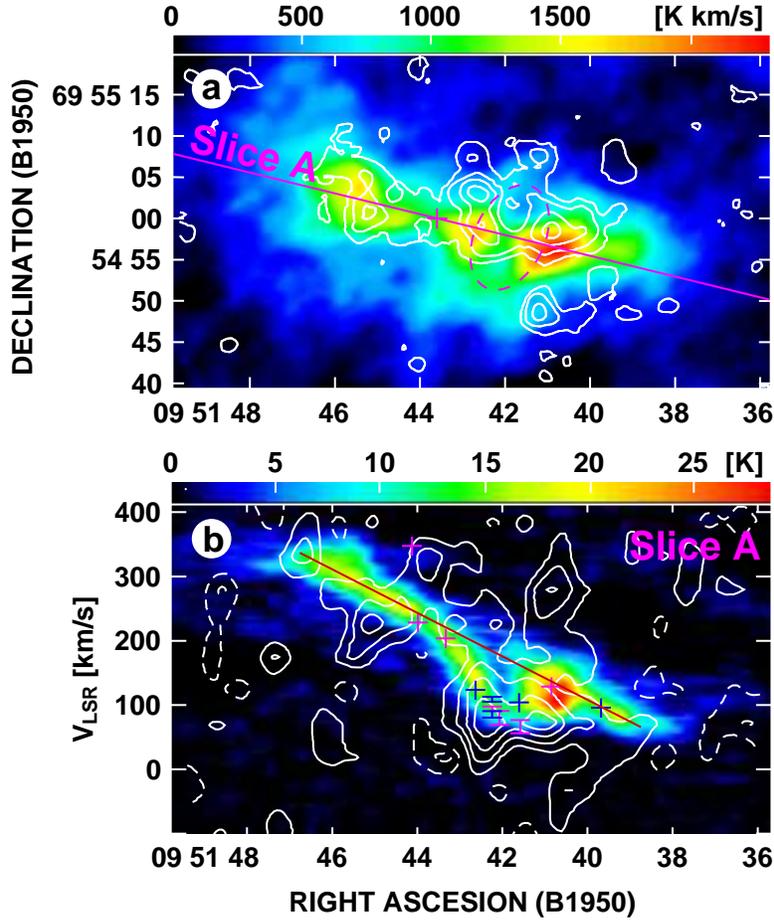}
\figcaption{
	(a) H41$\alpha$ integrated intensity image \citep[contours;][]{sea96}
	overlaid on the $^{12}$CO(1 -- 0) integrated intensity image
	\citep[colorscale;][]{mat00}.
	Values for the colorscale are indicated on the top of the figure.
	Dashed line (magenta) shows the brief structure of the molecular
	superbubble.
	Solid line (magenta) indicates the sliced region for
	a position--velocity (PV) diagram indicated below.
	The plus mark (magenta) is the same as Fig.~\ref{fig-uni-cont}.
	(b) PV diagram at the slice A.
	H41$\alpha$ image \citep[contours;][]{sea96} overlaid on
	the $^{12}$CO(1 -- 0) image \citep[colorscale;][]{mat00}.
	Values for the colorscale are indicated on the top of the figure.
	Solid line (red) indicates the rigid rotation velocity.
	Plus marks indicate OH \citep[magenta;][]{wel84,sea97} and
	H$_{2}$O \citep[blue;][]{bau96} masers with small velocity range.
	Bar marks indicate OH (magenta) and H$_{2}$O (blue) masers with
	velocity ranges indicated by the bars.
	H41$\alpha$ emission shows much larger deviation from the rigid
	rotation velocity than that of the molecular gas at the region of
	the superbubble
	(around $\alpha$(B1950)=$9^{\rm h}51^{\rm m}42^{\rm s}$).
	At this region, OH and H$_{2}$O masers are dynamically concentrated.
\label{fig-pv}}
\end{figure}

Molecular emission from the superbubble is strong in the southern part,
but weak in the northern part.
On the other hand, the images of atomic and ionized gas --- \ion{H}{1}
absorption line \citep{wil02}, radio recombination line
\citep[H41$\alpha$;][]{sea96} and forbidden line ([\ion{Ne}{2}];
\citealt{act95}) --- show stronger intensity from the northern part than from
the southern part \citep[Fig.~\ref{fig-pv}a; see also][]{wil99,wil02}.
The shell-like structure detected in the 408 MHz continuum by \citet{wil97}
clearly shows the complete superbubble structure.
These results seem to indicate that {\it the superbubble is not disrupted yet},
and the southern part of the superbubble is dominated by molecular gas and
the northern part by atomic and ionized gas.

The reason for this asymmetry is not clear.
One possibility is that the explosions, which made the superbubble, have
occurred preferentially north of the galactic disk.
This initial condition suggests that the southern part is rich in molecular
gas, but the northern part is not, so that the latter is more likely to be
dominated by atomic and ionized gas (i.e., less molecular gas and/or less
self-shielding of the molecular gas).
Indeed, the large-scale (kpc-scale) H$\alpha$ image of M82 shows that
the southern outflow is relatively confined to the minor axis of the galaxy,
but the northern outflow has a wider opening angle.
This difference suggests that the starburst is located slightly north of
the galactic disk, and the material distribution asymmetry causes the different
collimations in the northern and southern outflows \citep{sho98}.
Such asymmetry in the structure of bubbles can also be seen in other galaxies
(e.g., our Galaxy, \citealt*{nor96}; M101, \citealt*{kam91}),
although their low-density regions are thought to be already disrupted.

\subsubsection{Kinematics of Ionized Gas}
\label{dis-inducesb-ion}

We compared the position-velocity (PV) diagram of molecular gas with that
of a radio recombination line \citep[H41$\alpha$;][]{sea96} and found that
{\it the deviation from the rigid rotation of the recombination line is
larger than that of the molecular gas} (Fig.~\ref{fig-pv}b).
Similar deviation difference is also seen in the comparison between
the molecular gas and the [\ion{Ne}{2}] line \citep{wil99}.
Since the deviation is evidence of the superbubble expansion
\citep{nei98,wei99,wil99,mat00,wil02},
the ionized gas is kinematically related to the superbubble, but moves
faster than the neutral gas.

\subsubsection{Distributions and Kinematics of Masers}
\label{dis-inducesb-maser}

We compiled all the observed OH and H$_{2}$O masers toward M82
\citep*{wel84,bau96,sea97} and found that {\it the masers are obviously
concentrated at the 100 GHz continuum emitting regions, namely, the intense
starburst regions} (Fig.~\ref{fig-cont-co}).
In addition, we overplotted the masers on the PV diagram of molecular gas
and found that {\it the OH and H$_{2}$O masers are strongly concentrated at
the superbubble with good velocity coincidences} (Fig.~\ref{fig-pv}b).
Pumping mechanisms for OH masers are collisions ($\approx$ shocks) and/or
IR/UV radiation ($\approx$ radiation from \ion{H}{2} regions), and those
for H$_{2}$O masers are basically shocks \citep{eli92}.\footnotemark
\footnotetext{Strictly speaking, H$_{2}$O maser sources need to have
  (1) high H$_{2}$ number density of $\sim10^{9}$ cm$^{-3}$,
  (2) high temperature of $\sim400$ K, and (3) pumping sources.
  In our Galaxy, the maser-emitting sources are usually shocked molecular gas,
  but in the center of galaxies, dense molecular gas disks/tori around active
  galactic nuclei (AGNs) are also possible, such as H$_{2}$O megamasers toward
  the AGN of NGC 4258 \citep{miy95}.
  As mentioned below, since the luminosity of H$_{2}$O masers in M82 is
  similar to the Galactic sources, we think H$_{2}$O masers in M82 are coming
  from shocked molecular gas.}
Indeed, observations of our Galaxy indicate that these masers can be seen
around \ion{H}{2} regions \citep{gau87,eli92} and at the shock fronts of
supernova remnants \citep{ari99}.
Furthermore, masers in M82 are not so luminous as megamasers around AGNs,
but much more similar to the Galactic sources \citep{bau96,sea97}.
In addition, recent interferometric SiO line observations toward
the superbubble suggest that the SiO line may trace shocked regions, and
the shocked regions may locate the inner wall of the superbubble \citep{gar01},
consistent with our result.

\subsubsection{Distribution of Diffuse Hard X-ray Emission}
\label{dis-inducesb-xray}

\begin{figure}
\plotone{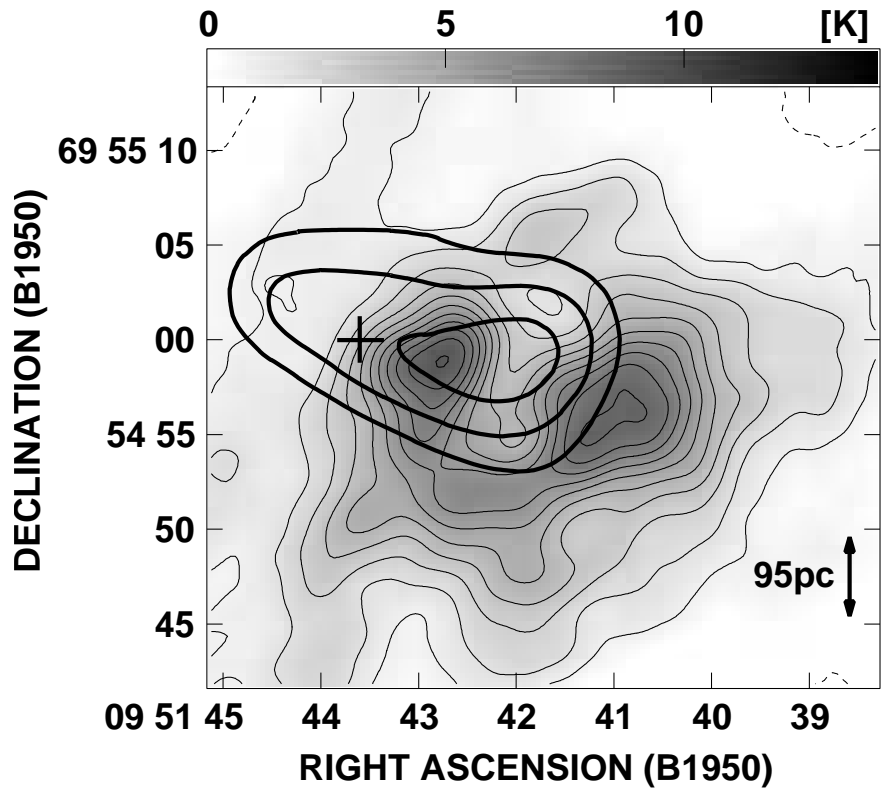}
\figcaption{
	Diffuse hard X-ray image \citep[thick contour map;][]{gri00}
	overlaid on the $^{12}$CO(1 -- 0) superbubble image
	\citep[thin contour with grayscale map;][]{mat00}.
	The plus mark is the same as Fig.~\ref{fig-uni-cont}.
	The linear scale at M82 is shown in the bottom-right of
	the figure.
\label{fig-co-xray}}
\end{figure}

Both inside and outside the superbubble, various X-ray sources can be seen from
soft/medium ($\sim0.1-2$ keV) to hard components ($\sim3-10$ keV).
The soft and medium components show extended structures, which are believed
to originate from hot thermal plasma \citep{tsur97,str97,gri00}.
The soft component corresponds well with the large-scale ionized gas outflows
\citep{wat84,kro85,str97,sho98}.
However, the medium component is much more concentrated \citep{tsur97}, and its
peak position is located at the center of the superbubble \citep{mat00,mat99}.
Most of the hard component, on the other hand, comes from point sources
\citep{tsur97}, which are thought to be neutron stars, stellar-mass black
holes, and an intermediate-mass black hole
\citep{mat99,pta99,mat00,mat01,kaa01}.
Some fraction of the hard component, however, comes from a diffuse source with
an extent of $7\farcs2\times5\farcs4$ \citep{gri00}.
Figure~\ref{fig-co-xray} shows the diffuse hard X-ray image
\citep[thick contour map;][]{gri00} overlaid on the the $^{12}$CO(1 -- 0)
superbubble image \citep[thin contour with greyscale map;][]{mat00}.
The diffuse hard X-ray emission is extended toward the center of
the galaxy, but the peak is located inside of the molecular superbubble.
Since the diffuse hard X-ray data do not have any line-of-sight information,
we do not know where is the actual location(s) of the emitting source(s), but
clearly there is some source toward the inner part of the superbubble.
This position correspondence suggests that some of the diffuse hard X-ray
component locates inside the superbubble.

This component could be due to hot thermal plasma \citep{gri00}, which is also
supported by 408 MHz observations \citep[free-free absorption has been detected
at the superbubble;][]{wil97}, but the possibility of inverse Compton
scattering cannot be ruled out \citep{gri00}.
Assuming that the diffuse hard X-ray component is caused by hot thermal plasma,
we can calculate the velocity of the plasma.
The sound speed, i.e., the velocity, of protons, $C_{\rm p}$, in the plasma
can be expressed as $C_{\rm p} \sim \sqrt{kT/m_{\rm p}}$, where $k$ is
the Boltzmann constant, $T$ is temperature (K), and $m_{\rm p}$ is
the proton mass.
Since the plasma temperature, $kT$, is estimated as $2.4-4.1$ keV
\citep{gri00}, the velocity can be calculated as
$C_{\rm p} \sim (5-6) \times 10^{2}$ km s$^{-1}$.
This number is $5-12$ times faster than the expansion velocity of the
superbubble, hence the superbubble should be closed and the plasma should be
overpressurized, if the plasma is in the superbubble.
Indeed, the superbubble seems to be closed (\S\ref{dis-inducesb-struct}), and
\citet{gri00} suggest that the plasma seems to be overpressurized.
In addition, the expansion velocity of the ionized gas\footnotemark
\footnotetext{In this paper, we used ``ionized gas'' as $10^{4}$ K gas that
	can be seen with recombination lines, and ``hot plasma'' as $10^{7}$ K
	gas that can be seen with thermal X-ray emission, to distinct these
	two kind of gas.}
in the superbubble
is faster than that of the molecular gas, and masers are concentrated at
the superbubble (\S\ref{dis-inducesb}).
These support the idea that the some of the diffuse hard X-ray component
is located inside the superbubble.

As mentioned above, the possibility that the diffuse hard X-ray component might
originate from inverse Compton scattering cannot be ruled out \citep{gri00}.
If true, it indicates that nonthermal high-energy electrons are strongly
concentrated inside the superbubble, because the diffuse hard X-ray emission
peaks at there.
The trapping mechanism of the nonthermal high-energy electrons is unclear,
but the magnetic field in and around the superbubble may be playing
an important role.

\subsubsection{Induced Starburst}
\label{dis-inducesb-inducesb}

In summary, the expanding superbubble seems to be still closed,
the overpressurized hot plasma may be filling inside of the superbubble,
ionized gas is spatially and kinematically correlated with the expansion of
superbubble but moves faster, the masers are also spatially and kinematically
located at the inner edge of the superbubble, and the free-free emission is
located at the inner edge of the superbubble.
The configuration of plasma, ionized gas, and superbubble is very similar
to that of the expanding supernova remnants in our Galaxy
\citep[e.g.,][]{whi91}; recombination lines that traces ionized gas emit
from shock-heated gas and the gas evaporate from the surrounding material
that forms shell structures.
The faster velocity than the surrounding molecular superbubble and the
existence of the masers support this idea.
These structural and kinematical features therefore suggest that
{\it the hot plasma expanded the superbubble, created the shocked regions at
the inner-edge of the superbubble, and induced the active starbursts at there.}

Sequential star formation in molecular clouds induced by expanding ionized gas
($\approx$ shocks) has been suggested theoretically
\citep[e.g.,][]{mcc87,elm77}.
In our Galaxy, this sequential star formation has also been suggested
observationally \citep*[e.g.,][]{sug89,sug95,yam99}.
In addition, these observational papers suggest that the induced star forming
regions tend to make more massive stars than isolated quiescent dark clouds.
We therefore conclude that the present starburst (massive star forming) regions
in M82 are induced in ways similar to those of the sequential star formation
regions in our Galaxy.

\begin{figure}
\plotone{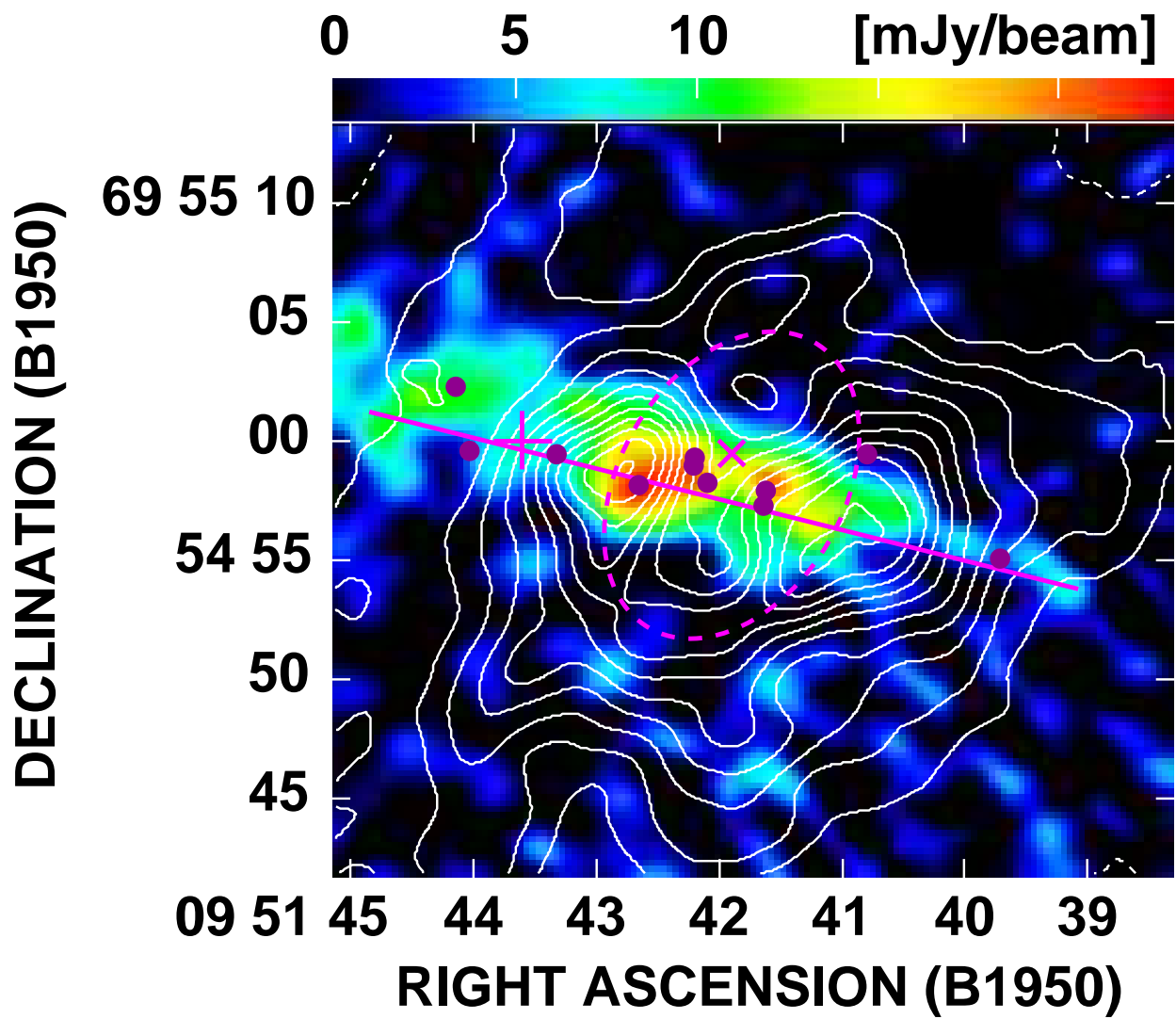}
\figcaption{
	$^{12}$CO(1 -- 0) image \citep[contours;][]{mat00}
	overlaid on 100 GHz continuum image (colorscale).
	Values for the colorscale are indicated on the top of the figure.
	The dashed line (magenta) illustrates the outline of the molecular
	superbubble.
	The solid line (magenta) indicates the sliced region for the
	intensity plots of the superbubble shown in Fig.~\ref{fig-slice}.
	The plus mark (magenta) is the same as Fig.~\ref{fig-uni-cont},
	and the cross mark (magenta) indicates the central position of
	the ``2.2 $\micron$ secondary peak'' \citep{die86}.
	The filled purple circles indicate the positions of the OH and H$_{2}$O
	masers \citep{wel84,bau96,sea97}.
\label{fig-cont-co}}
\end{figure}

At the center of the superbubble, on the other hand, there is almost no
molecular gas emission, neither diffuse ($^{12}$CO) nor dense (HCN), from both
spatial and dynamical points of view
\citep[Figs.~\ref{fig-cont-co}, \ref{fig-pv}b; see also][]{mat00}.
Furthermore, the 100 GHz continuum emission is also weak at this region
compared to the inner edge of the superbubble (see Fig.~\ref{fig-slice}).
Since stars are believed to be made from molecular clouds, especially from
dense parts \citep*[e.g.,][]{lad92,sol92,koh99}, and free-free (100 GHz
continuum) emission is produced by massive stars, these suggest that
the starburst at the center of the superbubble begins to cease, and
the number of massive stars is decreasing by supernova explosions.

From these results and discussions, we propose an evolution of starburst around
the superbubble in M82.
First, the energetic explosions occurred as a consequence of the localized
starburst at the center of the superbubble \citep{mat00}.
The resultant shock waves ionized the surrounding ISM, produced the hot thermal
plasma and ionized gas, and swept them outward.
The neutral ISM (e.g., molecular gas), had also been swept outward and produced
the expanding molecular superbubble.
Most of the molecular gas at the central starburst region was blown away and/or
ionized by the energetic explosions and/or by the strong UV radiation from
massive stars, so that the starburst begins to cease.
The expansion of the superbubble inside the galactic disk, which is rich in
ISM, had compressed the ISM and caused a concentration around the superbubble.
This concentration appears as double peaks by the edge-brightening effect
\citep{wil99}, and these double peaks correspond to the well known ``central
peak'' and the inner side of the ``SW lobe,'' which can be seen in molecular
gas images \citep[e.g.,][]{car91,she95}.
The inner edge of this concentration is the shock front of the superbubble
expansion, which caused the shocked regions and induced star forming regions
in the molecular gas, and produced the masers and free-free continuum emissions
at these regions.
These induced star forming regions correspond to the present starburst regions
in M82.
This result is the first clear evidence of self-induced starburst in
an external galaxy.

The size-scale of this self-induced starburst is rather small, about 200 pc,
and located off from the nucleus (see Fig.~\ref{fig-cont-co}).
This localized and offset self-induced starburst may be the cause of
inconsistent conclusions of inward/outward starburst propagations
(see \S\ref{intro}).
The detailed studies at various wavelengths with comparing the positions
of superbubbles \citep[e.g.,][]{wil02} will give us the detailed starburst
propagation and evolution mechanisms.

\subsection{Starburst Timescale}
\label{dis-timescale}

Recent observations and model calculations suggest that the central region of
M82 has experienced at least two starbursts within $\sim10^{7}$ years.
Optical emission line study suggests that the large-scale outflows originate
from at least two \citep{sho98}, maybe several \citep{mar98} different burst
regions.
Model calculations based on infrared spectroscopic observations required two
starbursts within $\sim10^{7}$ years.
The first starburst can be explained by the large-scale outflows and red
supergiants, and the second one by the UV flux \citep{rie93,for03}.
The molecular and dust outflows, which can be seen in both interferometric
\citep{mat00,wal02} and single-dish observations
\citep{nak87,kun97,alt99,tay01}, extend not only from the superbubble region,
but also everywhere from the molecular disk.

Our observations, on the other hand, show the past starburst, which has made
the superbubble
(expansion timescale of $\sim(1-2)\times10^{6}$ years; \S\ref{dis-mf-calc}),
and the present induced starburst, which is believed to be younger than
the expanding timescale of the superbubble (\S\ref{dis-mf}).
In addition, the superbubble still has closed structure
(\S\ref{dis-inducesb-struct} and \ref{dis-inducesb-xray}).
Furthermore, numerical simulations indicate that large-scale outflows are
created by energetic starbursts that can produce superbubbles
\citep[e.g.,][]{tom88}.

These observations and simulations indicate that the energetic starbursts in
M82, which produced the large-scale outflows, occurred at least twice in this
$\sim10^{7}$ years, and the recent starbursts (the starburst that made
the superbubble and the present starburst) occurred on an order-of-magnitude
shorter timescale.
These indications suggest that several starbursts had occurred in this
$\sim10^{7}$ years.
It is not clear whether starbursts occurred intermittently or continously
within $10^{6}$ years or shorter timescale.
However, if you average over long timescale, say order of $10^{7-8}$ years,
starburst in M82 seems occurring continuously, with the self-induced mechanism.
This self-induced starburst would continue till most of the molecular gas
is consumed by the star formation, blown away from the disk region of
the galaxy, and/or is dissociated by strong UV radiation, and this timescale
may correspond to the timescale of the whole starburst phenomena of
$10^{7-8}$ years \citep[e.g.,][]{rie88,hec98}.

These self-induced/self-regulating mechanisms would be very important for
other starburst galaxies.
Recent optical spectroscopic and X-ray observations show that most of starburst
galaxies have large-scale outflows \citep[e.g.,][]{leh96,mar98,dah98}.
Mid-infrared observations of starburst galaxies indicate that the timescale of
most starbursts are $10^{6}-10^{7}$ years \citep[e.g.,][]{tho00,van98}.
These results are consistent with those of M82, and strongly suggest that most
of the starbursts would have experienced similar starburst evolution as M82.

\section{CONCLUSIONS}
\label{conclude}

Our new high spatial resolution 100 GHz continuum (i.e., free-free emission)
image in the central region of M82 clearly shows that present starburst
regions are strongly concentrated at the inner edge of the expanding molecular
superbubble.
The Lyman continuum photon numbers from the starburst regions (i.e.,
the strength of the starburst) are an order of magnitude larger than those in
the active high-mass star forming regions in our Galaxy (NGC 3603 and W51)
and nearby galaxies (IC 342 and M51).

The internal structure of the present superbubble, from the center, shows
a red supergiant star cluster (``2.2 $\mu$m secondary peak''), overpressurized
hot thermal plasma or high energy electrons (diffuse hard X-ray emission),
high velocity ionized gas (H41$\alpha$ and [\ion{Ne}{2}]), shocked molecular
gas (masers and SiO emission) and the present starburst regions (free-free
emission), and the expanding molecular superbubble located furthermost outside.
This structure and the physical properties of these objects/mediums suggest
that the present starburst has been induced by the past starburst that made
the superbubble, namely, self-induced starburst.
The existence of kpc-scale outflows, a few hundred parsec-scale superbubble,
and induced starburst inside the superbubble suggest that the starbursts in
M82 have occurred several times in this $10^{6}-10^{7}$ years.
Since the appearance of the starburst (e.g., outflows) in M82 is typical in
other starburst galaxies, this starburst mechanism will be very important for
the studies of other starburst regions/galaxies and galaxy formation.

\acknowledgements

We would like to thank P. T. P. Ho, N. Fukuda, S. Ikeuchi, M. Saito,
K. Sugitani, and anonymous referee for helpful discussions and comments.
We also thank A. F. Omundsen for carefully reading our manuscript.
Thanks are also due to Y. Ohyama for providing the Subaru H$\alpha$
image for us.
We are grateful to the NRO staff for the operation and improvement of NMA.

\clearpage

\begin{deluxetable}{cccccccc}
\tabletypesize{\scriptsize}
\tablecaption{Observed and Calculated Properties of the Ongoing Starburst
	Regions. \label{tab1}}
\tablehead{\colhead{Peak\tablenotemark{a}} & \colhead{R.A.} & \colhead{Dec.}
	& \colhead{$F_{\rm T}$\tablenotemark{b}}
	& \multicolumn{2}{c}{$N_{\rm Ly}$\tablenotemark{c}}
	& \multicolumn{2}{c}{$N_{\rm > 8 M_{\odot}}$\tablenotemark{d}}\\
	\colhead{} & \colhead{$09^{\rm h}51^{\rm m}$} & \colhead{69\arcdeg}
	& \colhead{[mJy beam$^{-1}$]}
	& \multicolumn{2}{c}{[photons s$^{-1}$]} & \colhead{}\\
	\colhead{} & \colhead{(B1950)} & \colhead{(B1950)} & \colhead{}
	& \colhead{($T_{\rm e}$ = 5,000 K)}
	& \colhead{($T_{\rm e}$ = 10,000 K)}
	& \colhead{($T_{\rm e}$ = 5,000 K)}
	& \colhead{($T_{\rm e}$ = 10,000 K)}
	}
\startdata
A & 45\fs7 & 55\arcmin04\farcs1 & 13.2
	& $3.2\times10^{52}$ & $2.4\times10^{52}$
	& $\phantom{1}9.2\times10^{3}$ & $\phantom{1}6.7\times10^{3}$ \\
B & 44\fs4 & 55\arcmin01\farcs8 & 12.3
	& $3.0\times10^{52}$ & $2.2\times10^{52}$
	& $\phantom{1}8.6\times10^{3}$ & $\phantom{1}6.3\times10^{3}$ \\
C & 43\fs1 & 55\arcmin01\farcs3 & 12.6
	& $3.1\times10^{52}$ & $2.3\times10^{52}$
	& $\phantom{1}8.8\times10^{3}$ & $\phantom{1}6.5\times10^{3}$ \\
D & 42\fs6 & 54\arcmin58\farcs1 & 23.1
	& $5.7\times10^{52}$ & $4.2\times10^{52}$
	& $16.1\times10^{3}$ & $11.8\times10^{3}$ \\
E & 42\fs3 & 54\arcmin58\farcs6 & 21.2
	& $5.2\times10^{52}$ & $3.8\times10^{52}$
	& $14.7\times10^{3}$ & $10.8\times10^{3}$ \\
F & 41\fs6 & 54\arcmin58\farcs3 & 20.2
	& $5.0\times10^{52}$ & $3.6\times10^{52}$
	& $14.0\times10^{3}$ & $10.3\times10^{3}$ \\
G & 41\fs2 & 54\arcmin56\farcs6 & 15.3
	& $3.8\times10^{52}$ & $2.8\times10^{52}$
	& $10.7\times10^{3}$ & $\phantom{1}7.8\times10^{3}$ \\
\enddata
\tablenotetext{a}{Peak positions are indicated in Fig.~\ref{fig-uni-cont}.}
\tablenotetext{b}{100 GHz continuum flux at the peak positions.
	The beam size is $2\farcs3\times1\farcs9$
	\citep[43 pc $\times$ 36 pc at D = 3.9 pc;][]{sak99}.
	$1\sigma$ uncertainty is 2.0 mJy beam$^{-1}$.}
\tablenotetext{c}{Production rates of Lyman continuum photons calculated with
	$F_{\rm T}$ averaged over the beam size.
	$1\sigma$ uncertainties are $0.5\times10^{52}$ photons s$^{-1}$ for
	$T_{\rm e}$ = 5,000 K, and $0.4\times10^{52}$ photons s$^{-1}$ for
	$T_{\rm e}$ = 10,000 K. See text for details.}
\tablenotetext{d}{Number of stars with their masses larger than
	8 M$_{\odot}$, assuming a mass function.
	$1\sigma$ uncertainties are $1.4\times10^{3}$ for $T_{\rm e}$ = 5,000 K,
	and $1.0\times10^{3}$ for $T_{\rm e}$ = 10,000 K.
	See text for details.}
\end{deluxetable}

\clearpage

\clearpage

\clearpage

\clearpage

\clearpage

\clearpage

\end{document}